\documentclass[12pt]{iopart}
\newcommand{\bec}[1]{\mbox{\boldmath $ #1$}}
\usepackage{graphicx}
\begin{document}
\title[]{Growth rate of small-scale dynamo at low magnetic Prandtl numbers}
\author{Nathan Kleeorin and Igor Rogachevskii}
\address{Department of Mechanical Engineering,
Ben-Gurion University of the Negev, P.O.Box 653, Beer-Sheva 84105,  Israel; \\
NORDITA, AlbaNova University Center,
Roslagstullsbacken 23, SE-10691 Stockholm, Sweden}
\ead{gary@bgu.ac.il}

\date{\today}
\begin{abstract}
In this study we discuss two key issues related
to a small-scale dynamo  instability at low
magnetic Prandtl numbers and large magnetic
Reynolds numbers, namely: (i) the scaling for the
growth rate of small-scale dynamo instability in
the vicinity of the dynamo threshold; (ii) the
existence of the Golitsyn spectrum of magnetic
fluctuations in small-scale dynamos. There are
two different asymptotics for the small-scale
dynamo growth rate: in the vicinity of the
threshold of the excitation of the small-scale
dynamo instability, $\lambda \propto
\ln\left({\rm Rm} / {\rm Rm}^{\rm cr}\right)$,
and when the magnetic Reynolds number is much
larger than the threshold of the excitation of
the small-scale dynamo instability, $\lambda
\propto {\rm Rm}^{1/2}$, where ${\rm Rm}^{\rm
cr}$ is the small-scale dynamo instability
threshold in the magnetic Reynolds number ${\rm
Rm}$. We demonstrated that the existence of the
Golitsyn spectrum of magnetic fluctuations
requires a finite correlation time of the random
velocity field. On the other hand, the influence
of the Golitsyn spectrum on the small-scale
dynamo instability is minor. This is the reason
why it is so difficult to observe this spectrum
in direct numerical simulations for the
small-scale dynamo with low magnetic Prandtl
numbers.
\end{abstract}

\maketitle

\section{Introduction}

Generation of magnetic field by turbulent motions
of conducting fluid  is a fundamental mechanism
of magnetic fields observed in stars, galaxies
and planets. There are different kinds of
turbulent dynamos: large-scale and small-scale
dynamos. The large-scale mean-field dynamo
implies that the amplification of magnetic field
occurs at scales which are much larger than the
maximum scale of the turbulent motion. This kind
of dynamo includes: (i) the $\alpha \Omega$ and
$\alpha^2 \Omega$ dynamos caused by the combined
action of the $\alpha$ effect and differential
rotation (see, e.g.,
\cite{M78,P79,KR80,ZRS83,RSS88}); (ii) $\alpha^2$
dynamo in helical turbulence; and (iii) the shear
dynamos in non-helical turbulence
\cite{RK03,BR05,YSR08}.

On the other hand, generation of magnetic
fluctuations occurs at scales which are smaller
than the maximum scale of the turbulent motions
(see, e.g., reviews
\cite{ZMRS88,ZRS90,CG95,KA92,BS05,SIC07}).
Self-excitation of magnetic fluctuations with a
zero  mean magnetic field is called a small-scale
dynamo. The mechanisms of the small-scale dynamo
action are different depending on magnetic
Prandtl numbers ${\rm Pm} = \nu/\eta$, where
$\nu$ is the kinematic viscosity of the fluid and
$\eta$ is the magnetic diffusion due to
electrical conductivity of the fluid. For large
magnetic Prandtl numbers, the self-excitation of
magnetic fluctuations is caused by the random
stretching of the magnetic field by the smooth
velocity fluctuations (see, e.g.,
\cite{B50,MS63,ZRM84,ZMRS88,ZRS90,KR94,KRE94,S98,SSF12}).
This type of dynamo has been comprehensively
studied in direct numerical simulations (DNS) of
forced turbulence \cite{SCT04,HBD04,SHB05,FCS11}
and turbulent convection \cite{BR96,CAT99}. The
nature of small-scale dynamo for low magnetic
Prandtl numbers is different, e.g., it is driven
by the inertial-range velocity fluctuations at
the resistive scale. The small-scale dynamo at
low magnetic Prandtl numbers has been studied
analytically (see, e.g.,
\cite{K68,RK97,RK99,V02,BC04,CMV06,AH07}) for a
Gaussian white-noise velocity field (so called
the Kazantsev-Kraichnan model)  and numerically
(see, e.g., \cite{SIC07,ISC07,BR11}) in a number
of publications. Since the magnetic energy is not
conserved, the second moment of magnetic field
has anomalous scalings \cite{V96,RK97}.

The small-scale dynamo instability is excited
when the magnetic Reynolds number, ${\rm Rm}$, is
larger than the critical magnetic Reynolds
number, ${\rm Rm}^{\rm cr}$. Analytical models
based on the Kazantsev-Kraichnan model of a
homogeneous, isotropic, non-helical and
incompressible velocity field, yield ${\rm
Rm}^{\rm cr} \approx 410$ at very low magnetic
Prandtl numbers \cite{RK97}. Compressibility of
fluid flow causes strong increase of the critical
magnetic Reynolds number at ${\rm Pm} \ll 1$ (see
\cite{RK97}). Similar tendency also has been
recently demonstrated in analytical study
\cite{SSF12} at large Prandtl numbers. Direct
numerical simulations of small-scale dynamo in
\cite{SIC07,ISC07,BR11} of the Navier-Stokes
turbulence show that ${\rm Rm}^{\rm cr}$ is
around $200$ for small magnetic Prandtl numbers,
and it is at three times larger than for the
small-scale dynamo at large and moderate Prandtl
numbers (see \cite{SCT04,HBD04,SHB05}). These DNS
results at large and moderate Prandtl numbers are
in agreement with different analytical models
\cite{ZRS90,KR94,BC04,SSF12}.

The existence of the small-scale dynamo for a
large number of turbulent spectra at large
Prandtl numbers has been demonstrated in
\cite{SSF12}. When ${\rm Pm} \sim 1$ the
small-scale dynamo exists even in the regime of
very large Mach numbers (see the DNS results in
\cite{FCS11}). This study has also shown that,
for low Mach numbers $(\sim 0.1)$, the ratio of
the growth rate of turbulence driven by
solenoidal and compressive forcing is about $30$.
However, for higher Mach numbers ($\sim 10)$,
this ratio is about $2$.

The small-scale dynamo action is  different from
the turbulent induction effect that cases
production of the anisotropic magnetic
fluctuations by the tangling of the mean magnetic
field by the velocity fluctuations (see, e.g.,
\cite{G60,M61,RS82,KRR90,KRA94,KMR96}). This
effect cannot be described in terms of the
small-scale dynamo instability.

In spite of a number of studies of small-scale
dynamo at low magnetic Prandtl numbers, there are
some key questions  that are subject of
discussions in the literature. One of them is
related to the scaling for the growth rate
$\lambda$ of small-scale dynamo instability at
low magnetic Prandtl numbers in the vicinity of
the dynamo threshold. Our analysis performed in
this study and even numerical solution of the
dynamo equations for a Gaussian white-noise
velocity field obtained in \cite{MB10} imply that
there are two different asymptotics for the
dynamo instability growth rate: (i) in the
vicinity of the threshold of the excitation of
the small-scale dynamo instability and (ii) far
from the threshold of the small-scale dynamo
instability.

Other issue studied here is related to an
existence of the Golitsyn spectrum, $k^{-11/3}$,
of magnetic fluctuations \cite{G60,M61} in the
small-scale dynamo with low magnetic Prandtl
numbers. This spectrum of magnetic fluctuations
has been observed in the laboratory experiments
\cite{OPF98,BMP02}, in the large-eddy-simulations
\cite{PPP04} and in the direct numerical
simulation \cite{SIC07,BR11} of the small
magnetic Prandtl numbers magnetohydrodynamic
(MHD) turbulence. In the present study we discuss
conditions for the existence of the Golitsyn
spectrum.

The small-scale dynamo mechanism appears to be
responsible for the random magnetic fields in the
interstellar medium and in galaxy clusters
\cite{BBM96,BS05,SSH06,EV06,SIC07}. A number of
studies recently pointed out also the relevance
of the small-scale dynamo to amplify small seed
fields in galaxies and the intergalactic medium
(see, e.g.,
\cite{RKC08,ABK09,SBS10,SSB10,TOA12}). In
particular, DNS study in \cite{SSB10}
demonstrated that in the presence of turbulence,
weak seed magnetic fields are amplified by the
small-scale dynamo during the formation of the
first stars. Strong magnetic fields are generated
during the birth of the first stars in the
universe, potentially modifying the mass
distribution of these stars and influencing the
subsequent cosmic evolution (see \cite{SSB10}).
It was also noted in \cite{SBS10} that the
small-scale dynamo is very efficient during the
formation of the first stars and galaxies. During
gravitational collapse, turbulence is created
from accretion shocks, which may act to amplify
weak magnetic fields in the protostellar cloud.
Such turbulence is sub-sonic in the first
star-forming minihalos, and highly supersonic in
the first galaxies. It was concluded in
\cite{SBS10} that magnetic fields are
significantly enhanced before the formation of a
protostellar disk, where they may change the
fragmentation properties of the gas and the
accretion rate.

\section{Governing equations}

Let us study magnetic fluctuations with a zero
mean magnetic field at low magnetic Prandtl
numbers. In sections 2-3 we use the
Kazantsev-Kraichnan model \cite{K68} of the
$\delta$-correlated-in-time random  velocity
field. Using this model allows to get the
analytical results for the growth rate of the
small-scale dynamo instability. The results
remain valid also for the velocity field with a
finite correlation time if the second-order
correlation functions of the magnetic field vary
slowly in comparison to the correlation time of
the turbulent velocity field (see, e.g.,
\cite{ZRS90,KRS02}). The two-point instantaneous
correlation function of the magnetic field can be
presented in the form
\begin{eqnarray}
\langle b_i(t,{\bf x}) b_j(t,{\bf y}) \rangle = \tilde W(t,r)
\delta_{ij} + {r \tilde W'\over 2}  \, (\delta_{ij} - r_{ij}) ,
\label{A11}
\end{eqnarray}
where $\tilde W(t,r) = \langle b_{r}(t,{\bf x})
\, b_{r}(t,{\bf y}) \rangle$  is the longitudinal
correlation function, $b_r$ is the component of
magnetic field ${\bf b}$ in the direction ${\bf
r} = {\bf x} - {\bf y}$, $\, r_{ij} = r_i r_j /
r^2$ and $\tilde W' = \partial\tilde W/\partial
r$. This form of the second moment~(\ref{A11})
corresponds to the condition $\bec{\nabla} \cdot
{\bf b} = 0$
and an assumption of the homogeneous
and isotropic magnetic fluctuations. Equation for
the function $\tilde W(r,t)$ derived in the
framework of the Kazantsev-Kraichnan model of a
homogeneous, isotropic, non-helical,
incompressible and Gaussian white-noise velocity
field, reads
\begin{eqnarray}
{\partial \tilde W(t,r) \over \partial t} = {1 \over m(r)} \tilde W'' + \mu(r) \tilde W'
- {\kappa(r) \over m(r)} \tilde W ,
\label{B1}
\end{eqnarray}
(see \cite{K68,RK97}), where
\begin{eqnarray*}
{1 \over m(r)} &=& {2 \over {\rm Rm}} + {2 \over 3} \, [1 - F(r)],
\quad
\mu(r) = {4 \over m r} + \biggl({1 \over m} \biggr)' ,
\\
\kappa(r) &=& {2 m \over r} \, f'(r) ,
\quad
f(r) = F(r) + r F' / 3 ,
\end{eqnarray*}
and ${\rm Rm}=u_0 \, \ell_0/ \eta \gg 1$ is the
magnetic Reynolds  number, $u_0$ is the
characteristic turbulent velocity in the integral
scale $\ell_0$ and $ F' = dF(r) / dr$. Hereafter
equations are written in dimensionless variables:
length and velocity are measured in units of
$\ell_0$ and $u_0$. For a homogeneous, isotropic
and non-helical (with zero mean helicity) and
incompressible turbulent fluid velocity field the
correlation function $\langle \tau u_i({\bf x})
u_j({\bf x} + {\bf r}) \rangle$ is given by
\begin{eqnarray}
\langle \tau u_i({\bf x}) u_j({\bf y}) \rangle = {1 \over
3} \, \biggl[F(r) \, \delta_{ij} + {r F'\over 2}  \, (\delta_{ij} -
r_{ij}) \biggr] .
\label{B2}
\end{eqnarray}
The form of the continues function $F(r)$ with
different scalings in different ranges of scales
is constructed using the following reasoning. The
function $F(r)=1 - \sqrt{\rm Re} \, r^2$ is in
the viscous range of scales, $0 \leq r \leq
\ell_\nu/\ell_0$, while the function $F(r)=1 -
r^{4/3}$ is in the inertial range of scales,
$\ell_\nu/\ell_0 < r < 1$. At the boundary of
these ranges, $r=\ell_\nu/\ell_0$, these
functions coincide, where $\ell_\nu = \ell_0/{\rm
Re}^{3/4}$ is the  viscous scale and $\ell_0$ is
the integral scale of turbulence.

The solution of Eq. (\ref{B1}) can be obtained
using an asymptotic  analysis (see, e.g.,
\cite{ZRS90,RK97,RK99}). This analysis is based
on the separation of scales. In particular, the
solutions of Eq. (\ref{B1}) with a variable mass
have different regions with different functions
$m(r)$, $\, \mu(r)$ and $\kappa(r)$. Solutions in
these different regions and their derivatives can
be matched at their boundaries. The results
obtained by this asymptotic analysis are
presented below.

\section{Asymptotic behaviour of the growth rate of magnetic fluctuations}

Let us discuss the asymptotic behaviour of the
growth rate of magnetic  fluctuations with a zero
mean for small magnetic Prandtl numbers. We seek
for a solution of Eq.~(\ref{B1}) for the
longitudinal correlation function of the magnetic
field in the form: $\tilde W(t,r) = \exp (\lambda
\, t) \, W(r)$. In the viscous range of scales,
$0 \leq r \leq \ell_\nu/\ell_0$, the function
$F(r)=1 - \sqrt{\rm Re} \, r^2$ and the equation
for the function $W(r)$ is given by:
\begin{eqnarray}
r \, W'' + 4 \, W' + {10 \over 3} \, {\rm Pr}_m \, {\rm Re}^{3/2}\, r \, W = 0 ,
\label{C1}
\end{eqnarray}
where $W' = dW(r)/dr$. The solution of
Eq.~(\ref{C1}) is given by
\begin{eqnarray}
W(r) &=& r^{-3/2} J_{3/2} \left(\sqrt{10 \, {\rm
Pr}_m\over 3} \, {\rm Re}^{3/4} \, r \right)
\approx 1 - {{\rm Pr}_m \, {\rm Re}^{3/2} \over
3} \, r^2 , \label{C5}
\end{eqnarray}
(see \cite{RK97}), where $J_\alpha(y)$ is the
Bessel function of the first kind, and we have
taken into account that $W(r=0)=1$.

In the inertial range of scales, $\ell_\nu/\ell_0
< r < 1$, the  function $F(r)=1 - r^{4/3}$ and
the equation for the function $W(r)$ is given by:
\begin{eqnarray}
\left(1 + {1\over 3} \, {\rm Rm} \,
r^{4/3}\right) \, W'' &+& {4 \over r} \,\left(1+
{4\over 9} \, {\rm Rm} \, r^{4/3}\right) \, W'
\nonumber\\
&+& {\rm Rm} \, \left({52\over 27} \, r^{-2/3} -
{\lambda \over 2}\right) \, W = 0 , \label{B0}
\end{eqnarray}
where $\lambda$ is the growth rate of small-scale
dynamo instability. In the range of scales,
$\ell_\nu/\ell_0 < r \ll \ell_\eta/\ell_0$ the
equation for the function $W(r)$ reads:
\begin{eqnarray}
r^2 \, W'' + 4\, r \, W' + {52\over 27} \, {\rm
Rm} \, r^{4/3} \, W = 0, \label{B37}
\end{eqnarray}
where $\ell_\eta = \ell_0/{\rm Rm}^{3/4}$ is the
resistive  scale. The solution of Eq.~(\ref{B37})
is given by
\begin{eqnarray}
W(r) &=& r^{-3/2} J_{9/4} \left(\sqrt{13 \, {\rm
Rm} \over 3} \, r^{2/3} \right) \approx 1 - {{\rm
Rm} \over 3} \, r^{4/3} , \label{B38}
\end{eqnarray}
(see \cite{RK97}). On the other hand, in the
range of scales, $\ell_\eta/\ell_0 \ll r < 1$ the
equation for the function $W(r)$ is given by:
\begin{eqnarray}
9 \, r^2 \, W'' + 48 \, r \, W' + \left(52 - {27
\, \lambda \over 2} \, r^{2/3} \right) \, W = 0 .
\label{B3}
\end{eqnarray}
The solution of Eq.~(\ref{B3}) is
\begin{eqnarray}
W(r) = C \, r^{-13/6} K_{\alpha} \left(\sqrt{27
\,\lambda \over 2} \, \, r^{1/3} \right) ,
\label{C2}
\end{eqnarray}
(see  \cite{AH07}), where $K_{\alpha}(y)$ is the
real part of the modified Bessel  function
(Macdonald function) with $\alpha = (i/2) \,
\sqrt{39}$. This solution is chosen to be finite
at large $r$, with positively defined spectrum,
and it has the following asymptotics at scales $r
\ll \lambda^{-3/2}$ (see \cite{RK97}):
\begin{eqnarray}
W(r) = A_1 \, r^{-13/6} \, \cos\left(\sqrt{13
\over 12} \, \ln r + \varphi_0 \right) ,
\label{B5}
\end{eqnarray}
and at scales $\lambda^{-3/2} \ll r \ll 1$ (see
\cite{BC04}):
\begin{eqnarray}
W(r) = A_2 \, r^{-7/3} \exp \left(-\sqrt{27
\,\lambda \over 2} \, \, r^{1/3} \right) .
\label{B6}
\end{eqnarray}
Here $A_1$ and $A_2$ are the constants which are
proportional to the constant $C$.

In the range of scales $r \gg 1$, the turbulence
is absent $(F \to 0)$, $1 /m = 2 / 3$, $\mu(r)
=4/mr$ and
\begin{eqnarray}
W(r) = A_3 \, r^{-2}\,  \left(\sqrt{\lambda} +
r^{-1} \right) \, \exp(- \lambda r) , \label{B40}
\end{eqnarray}
(see \cite{RK97}), where $A_3$ is a constant.

The scaling for the growth rate of small-scale
dynamo instability which is far from the
threshold, is estimated as inverse resistive time
scale:
\begin{eqnarray}
\lambda \sim {u_\eta \over \ell_\eta} \sim {u_0 \over \ell_0} \, {\rm Rm}^{1/2} ,
\label{C3}
\end{eqnarray}
(see \cite{M61}), where $u_\eta= (\varepsilon \,
\ell_\eta)^{1/3}$ is the characteristic turbulent
velocity at the resistive scale, $u_0=
(\varepsilon \, \ell_0)^{1/3}$ and $\varepsilon$
is the dissipation rate of turbulent kinetic
energy. For the scaling $\lambda \propto {\rm
Rm}^{1/2}$, the condition $\lambda^{1/2} \,
r^{1/3} \gg 1$ implies $r \gg {\rm Rm}^{-3/4}$.
Note that the matching of the
solutions~(\ref{B38}), (\ref{B6}) and their
derivatives at the boundary of their regions
yields the dynamo growth rate~(\ref{C3}).

\begin{figure}
\centering
\includegraphics[width=14cm]{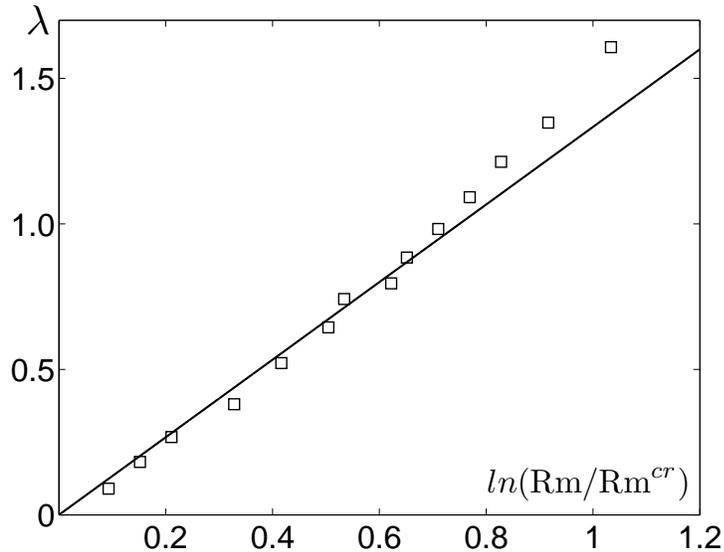}
\caption{\label{Fig1} The growth rate of
small-scale dynamo instability versus
$\ln\left({\rm Rm}/{\rm Rm}^{\rm cr}\right)$ in
the vicinity of  the instability threshold: solid
line corresponds to the scaling $\lambda=\beta \,
\ln\left({\rm Rm} / {\rm Rm}^{\rm cr}\right)$ and
squares are the results of the numerical solution
of the dynamo equation for $W(r)$ for the
Kazantsev-Kraichnan model of velocity field with
zero kinetic helicity in the inertial range of
scales taken from Fig.~1 in \cite{MB10}.}
\end{figure}

However, the scaling, $\lambda \propto {\rm
Rm}^{1/2}$, is not valid in  the vicinity of the
threshold of the dynamo instability. Indeed, in
the vicinity of the threshold when $\lambda \to
0$, there is only one range of the solution of
Eq.~(\ref{B3}), i.e., $\lambda^{1/2} \, r^{1/3}
\ll 1$. In this range of scales the solution of
Eq.~(\ref{B3}) is determined by Eq.~(\ref{B5}).
The matching of the solutions~(\ref{B38}),
(\ref{B5}) and~(\ref{B40}) and their derivatives
at the boundary their regions yields the
following growth rate of the small-scale dynamo
instability in the vicinity of the threshold:
\begin{eqnarray}
\lambda=\beta \, \ln\left({{\rm Rm} \over {\rm Rm}^{\rm cr}}\right) ,
\label{C4}
\end{eqnarray}
(see Appendix A), where $\beta=4/3$ is the
exponent of the scaling  of the correlation
function $F(r)$ (i.e., it is the exponent of the
turbulent diffusivity scaling). In Fig.~1 we plot
the growth rate~(\ref{C4}) of small-scale dynamo
instability versus $\ln\left({\rm Rm}/{\rm
Rm}^{\rm cr}\right)$ in the vicinity of the
threshold of small-scale dynamo instability. In
the same figure we also show the numerical
solution (squares) of the dynamo
equation~(\ref{B1}) performed in \cite{MB10} (for
the Kazantsev-Kraichnan model in inertial range
of scales of fluid motions for ${\rm Re} \sim
10^8$, and for ${\rm Rm} \leq 10^7$), which
demonstrates perfect agreement between the
scaling~(\ref{C4}), shown by solid line, and the
numerical solution of the dynamo equation.

Note that the solution of Eq.~(\ref{B3})
determined by Eq.~(\ref{B5}), is generally a fast
oscillating function at $r \ll 1$. However, for
the first mode with the maximum growth rate, the
spectrum of the eigenfunction is positively
defined (see \cite{ZRS90}). Since this solution
is only valid in the vicinity of the dynamo
threshold, the second and higher modes are not
excited. Therefore, the resulting spectrum of the
eigenfunctions are always positively defined.

In the present study we discuss only the regime
of small magnetic Prandtl numbers. In the case of
large magnetic Prandtl numbers and large fluid
Reynolds numbers the dynamo growth rate far from
the threshold is $\lambda \sim Re^{1/2}$, i.e.,
it is determined by the Kolmogorov time scale
(see, e.g., \cite{M61,S98,SSF12}). On the other
hand, for small magnetic Prandtl numbers the
dynamo growth rate far from the threshold is
determined by the resistive time scale.

\section{Magnetic fluctuations with the Golitsyn spectrum}

In this Section we study effect of magnetic
fluctuations with the Golitsyn  spectrum,
$k^{-11/3}$, \cite{G60,M61} on the small-scale
dynamo with low magnetic Prandtl numbers. This
spectrum can exist at the scales $\ell_\nu \leq r
\leq \ell_\eta$. Our goal is to determine the
longitudinal correlation function $W(r)$ that
corresponds to the Golitsyn spectrum. To this end
we use the induction equation for the
instantaneous magnetic field ${\bf H}(t,{\bf x})$
in an incompressible velocity field ${\bf
v}(t,{\bf x})$:
\begin{eqnarray}
{\partial {\bf H} \over \partial t} + ({\bf v} \cdot \bec{\nabla})
{\bf H} = ({\bf H} \cdot \bec{\nabla}) {\bf v} + \eta \Delta {\bf
H} .
\label{A1}
\end{eqnarray}
We seek for the solution of Eq.~(\ref{A1}) in the following form:
\begin{eqnarray}
{\bf H}(t,{\bf x}) = [{\bf B}(t)+{\bf b}(t,{\bf x})] \, \exp
(\lambda \, t/2) ,
\label{A2}
\end{eqnarray}
where ${\bf B}(t)$ is the magnetic field in the
scales which are much larger than the resistive
scale $\ell_\eta$, while ${\bf b}(t,{\bf x})$ is
the magnetic field in the scales which are
smaller than $\ell_\eta$. We consider the
magnetic dynamo regime, so that the total
magnetic field ${\bf H}$ grows in time
exponentially with the growth rate $\lambda/2$.
Now we average Eq.~(\ref{A1}) over the ensemble
of fluctuations generated in the scales
$\ell_\eta \ll \ell \ll \ell_0$, and subtract the
obtained averaged equation from Eq.~(\ref{A1}).
This yields equation for the magnetic field ${\bf
b}(t,{\bf x})$:
\begin{eqnarray}
{\partial {\bf b} \over \partial t} = ({\bf B} \cdot \bec{\nabla})
{\bf u} + (\eta \Delta - \lambda/2) \, {\bf b}  + {\bf
b}^N ,
\label{A3}
\end{eqnarray}
where ${\bf v}(t,{\bf x}) = {\bf V}(t)+{\bf
u}(t,{\bf x})$, $\, {\bf V}(t)$ is the velocity
field in the scales which are much larger than
the resistive scale $\ell_\eta$, while ${\bf
u}(t,{\bf x})$ is the velocity field in the
scales which are smaller than the resistive scale
$\ell_\eta$, and ${\bf b}^N= \bec{\nabla} \times
({\bf u} \times {\bf b} - \langle {\bf u} \times
{\bf b} \rangle)$  are the nonlinear terms.
Equation~(\ref{A3}) is written in the frame
moving with the velocity ${\bf V}(t)$. Using
Eq.~(\ref{A3}) and the momentum equation for the
velocity ${\bf u}(t,{\bf x})$ we derive equations
for the second moments of magnetic field
$h_{ij}({\bf k}) = \langle b_i({\bf k}) \,
b_j(-{\bf k}) \rangle$ and the cross helicity
tensor $g_{ij}({\bf k}) = \langle b_i({\bf k}) \,
u_j(-{\bf k}) \rangle$:
\begin{eqnarray}
{\partial h_{ij}({\bf k}) \over \partial t} &=& - i ({\bf B} {\bf
\cdot} {\bf k}) \, [g_{ij}({\bf k}) - g_{ji}(-{\bf k})]
- (\eta k^2 + \lambda/2) \, h_{ij} + h_{ij}^N ,
\label{A4} \\
{\partial g_{ij}({\bf k }) \over \partial t} &=&
i ({\bf B} {\bf \cdot} {\bf k}) \, \tilde
f_{ij}({\bf k}) - (\eta k^2 + \lambda/2) \,
g_{ij} + g_{ij}^N , \label{A5}
\end{eqnarray}
where $\tilde f_{ij}({\bf k}) = \langle u_i({\bf
k}) \, u_j(-{\bf k}) \rangle$, $\, h_{ij}^N =
\langle b^N_i ({\bf k}) b_j(-{\bf k}) \rangle +
\langle b_i({\bf k}) b^N_j(-{\bf k}) \rangle$ and
$g_{ij}^N = - i({\bf k} {\bf \cdot} {\bf B}) \,
h_{ij}({\bf k}) + \langle b^N_i ({\bf k})
u_j(-{\bf k}) \rangle + \langle b_i({\bf k})
u^N_j(-{\bf k}) \rangle$. Here $u^N_i$ are the
nonlinear terms in the momentum equation. Since
we have already taken into account the
exponential growth of the total magnetic field
${\bf H}$, we can drop the time derivatives in
Eqs.~(\ref{A4}) and~(\ref{A5}), because the
characteristic times of the variations of the
correlation functions $h_{ij}$ and $g_{ij}$ are
much larger than the time $\lambda^{-1}$.   Since
we describe magnetic fluctuations in the spatial
scales which are smaller than the resistive scale
$\ell_\eta$, we may drop the nonlinear terms
$h_{ij}^N$ and $g_{ij}^N$ in Eqs.~(\ref{A4})
and~(\ref{A5}) in the case of large magnetic
Reynolds numbers and low Prandtl numbers, because
they are small in these scales. Therefore, the
Eqs.~(\ref{A4}) and~(\ref{A5}) yield:
\begin{eqnarray}
(\eta \, k^2 + \lambda/2)^2 \, \langle b_i({\bf k}) \, b_j(-{\bf k}) \rangle
=2({\bf B} {\bf \cdot} {\bf k})^2 \, \langle u_i({\bf k}) \, u_j(-{\bf k}) \rangle .
\label{A6}
\end{eqnarray}
Now we introduce the normalized two-point correlation function
$w(r)$ of the magnetic field which is defined as follows:
\begin{eqnarray}
w(r) = {1 \over \langle B^2 \rangle} \, \langle H_m({\bf x}) \,
H_m({\bf y}) \rangle
= 1 +  {1 \over \langle B^2 \rangle} \, \langle b_m({\bf x}) \,
b_m({\bf y}) \rangle ,
\label{A7}
\end{eqnarray}
where $w' = dw(r) / dr$ and $r = |{\bf x}-{\bf
y}|$. We rewrite Eq.~(\ref{A6}) in ${\bf r}$
space using the inverse  Fourier transformation
(i.e., we use the following transformation $i k_i
\to \nabla_i$). This yields the following
equation for the normalized two-point correlation
function of the magnetic field:
\begin{eqnarray}
\biggl({d^2 \over dr^2} + {2 \over r} \, {d \over
dr} - a^2\biggr)^2 \, w(r) = - {2 \over 3} \,
{\rm Rm}^2 \, \left( {d^2 \tilde f(r) \over dr^2}
+ {2 \over r} {d \tilde f(r) \over dr} \right) +
3 \, a^4 , \label{A8}
\end{eqnarray}
where $a=(\lambda \, {\rm Rm}/2)^{1/2}$, $\,
\tilde f(r)=\tilde f_{mm}(r)$ and $\tilde
f_{ij}(r)$ is the two-point correlation function
of the velocity fluctuations written in ${\bf r}$
space. Equation~(\ref{A8}) is written in
dimensionless variables: length and velocity are
measured in units of $\ell_0$ and $u_0$, the
growth rate of the magnetic fluctuations
$\lambda$ is  measured in units of $u_0/\ell_0$
and the magnetic field is measured in units of
$B_0$. We also take into account that $\langle
({\bf B} \cdot \bec{\nabla})^2 \rangle \tilde f =
(1/3) \, (\tilde f'' + 2 \tilde f'/r)$.

Solution of Eq.~(\ref{A8}) which satisfies the following
boundary conditions: $w(r=0) = 3$ and $w'(r=0) = w'''(r=0) = 0$,
reads
\begin{eqnarray}
w(r) &=& 3 - {3 \, C_1 \over 4 \, a^3 \, r} \, [(a r - 1) \, \exp(a r) + (a
r + 1) \, \exp(- a r)]
\nonumber \\
&& + {55 \, {\rm Rm}^2 \over 2 \, (3^6) \,
a^{11/3} \, r} \, \biggl[(3 \, a r - 5) \, \exp(a
r) \, \gamma(2/3, a r)
\nonumber \\
&& + (3 \, a r + 5) \, \exp(-a r) \, \gamma(2/3,
-a r) \biggr], \label{A9}
\end{eqnarray}
where $w' = dw(r) / dr$, $\, C_1$ is a free constant and
$\gamma(\beta, x) = \beta^{-1} \, x^\beta \, \exp(-x) \,
M(1,1+\beta,x)$ is the incomplete gamma function which is related
to the confluent hypergeometric function $M(a,b,x)$. When $a r \ll
1$, Eq.~(\ref{A9}) for the two-point correlation function $w(r)$
is given by
\begin{eqnarray}
w(r) = 3 - C_1 \, r^2 + {1 \over 12} \, {\rm
Rm}^2 \, r^{8/3} \, \biggl[1 + {9\over 238} \,
{\rm Rm} \, \lambda \, r^2 \biggr]. \label{A10}
\end{eqnarray}
The function $w(r)$ is related to the longitudinal correlation
function $W(r)$, i.e., $w(r)=3 W(r) + r W'(r)$.
Equation~(\ref{A10}) rewritten for the longitudinal
correlation function $W(r)$, reads
\begin{eqnarray}
W(r) = 1 - {{\rm Rm} \, {\rm Re}^{1/2} \over 3}
\, r^2 + {1 \over 68} \, {\rm Rm}^2 \, r^{8/3} \,
\biggl[1 + {9\over 322} \, {\rm Rm} \, \lambda \,
r^2 \biggr], \label{A10}
\end{eqnarray}
where $\ell_\nu/\ell_0 \leq r \leq
\ell_\eta/\ell_0$, the  constant $C_1 \approx
(5/3) \, {\rm Rm} \, {\rm Re}^{1/2}$ is
determined by the matching of functions $W(r)$
determined by Eqs.~(\ref{C5}) and~(\ref{A10}) at
the point $r=\ell_\nu/\ell_0$. The scaling $W(r)
\propto {\rm Rm}^2 \, r^{8/3}$ corresponds to the
Golitsyn spectrum $M(k) \propto B_0^2 \,
\eta^{-2} \, \varepsilon^{2/3} \, k^{-11/3}$
\cite{G60}, where $\varepsilon$ is the rate of
dissipation of the turbulent kinetic energy and
$M(k) = (2/\pi) \int_0^{\infty} k r \sin (kr) \,
w(r) \, dr$. It follows from the latter equation
that the exponent $q$ in the spectrum function
$M(k) \propto k^{-q}$ and the exponent $p$ in the
scaling $W(r) \propto r^{p}$ are related as
follows $q=p+1$.

For small Prandtl numbers the constant $\tilde
C_1 = {\rm Rm} \, {\rm Re}^{1/2} / 3$ is larger
than $\tilde C_2 = {\rm Rm}^2 / 68$, and since $r
\ll 1$, the first and second terms in the right
hand side of Eq.~(\ref{A10}) dominate the
behavior of $W(r)$. This estimate implies that
the third term  in the right hand side of
Eq.~(\ref{A10}) resembling the Golitsyn spectrum
is negligible. This is the reason that the
influence of the Golitsyn spectrum on the
small-scale dynamo instability is minor. That is
why it is so difficult to observe the Golitsyn
spectrum in direct numerical simulations (DNS)
for the small-scale dynamo with low magnetic
Prandtl numbers.

Note that the existence of the Golitsyn spectrum
of magnetic fluctuations requires a finite
correlation time of the random velocity field,
i.e., the solution for  the small-scale dynamo
with the Golitsyn spectrum does not exist in the
framework of the Krachnan-Kazantsev model of the
delta-correlated-in-time turbulent velocity field
(see Appendix B). Indeed, for the derivation of
Eq.~(\ref{A8}) we did not used assumption about
the delta-correlated-in-time turbulent velocity
field. One of the indications of the finite
correlation time of random velocity field is
already seen in Eq.~(\ref{A8}), where the
high-order spatial derivatives arise. The
Kazantsev-Kraichnan model yields the dynamo
equation with spatial derivatives not higher than
the second-order spatial derivatives. On the
other hand,  it is well-known that even small yet
finite correlation time of random velocity field
causes the appearance of the higher-order spatial
derivatives in the dynamo equations (see, e.g.,
\cite{ZRS90,KRS02}).

The requirement of the finite correlation time of
random velocity field for the correct description
of the tangling magnetic fluctuations which have
the Golitsyn spectrum even follows from the
dimensional arguments. Indeed, the main balance
in the induction equation for the magnetic
fluctuations $({\bf B} \cdot \bec{\nabla}) {\bf
u} \sim D \Delta {\bf b}$ which yields the
Golitsyn spectrum, can be rewritten in the
following form: $\langle {\bf b}^2 \rangle \sim
\tau^2(\ell) \, [\langle {\bf u}^2 \rangle]^2 \,
{\bf B}^2 / D^2$. The latter equation implies the
requirement of the finite correlation time of
random velocity field for the correct description
of the tangling magnetic fluctuations. The
similar arguments are also valid for the $k^{-1}$
spectrum of magnetic fluctuations generated by
the tangling mechanism at low magnetic Prandtl
numbers in the scales $\ell_\eta \ll \ell \ll
\ell_0$  (see \cite{RS82,KRR90,KRA94,KMR96}).

We stress again that both magnetic fields, ${\bf
B}(t)$ and ${\bf b}(t,{\bf x})$, are small-scale
fields (in scales which are much less than the
integral scale $\ell_0$ of turbulence). In
particular, ${\bf B}(t)$ is the magnetic field in
the scales $\ell_\eta \ll \ell \ll \ell_0$, while
${\bf b}(t,{\bf x})$ is the magnetic field in the
scales $\ell_\nu \ll \ell \ll \ell_\eta$. These
fields belong to the same mode generated by the
same small-scale dynamo mechanism. In this
section we used two magnetic fields, ${\bf B}(t)$
and ${\bf b}(t,{\bf x})$, to describe interaction
of the magnetic fields of different scales by the
tangling of the field ${\bf B}(t)$ of the
velocity fluctuations which produces additional
anisotropic magnetic fluctuations with the
Golitsyn spectrum. The latter mechanism is the
turbulent magnetic induction that is different
from the small-scale dynamo.

\section{Conclusions}

In this study we investigated some key issues of
small-scale dynamos in random velocity field with
large fluid Reynolds numbers,  a zero mean
magnetic field and low magnetic Prandtl numbers.
Contrary to the claim in \cite{MB10}, there are
two different asymptotics for the dynamo growth
rate: in the vicinity of the threshold of the
excitation of the dynamo instability [$\lambda
\propto  \ln\left({\rm Rm} / {\rm Rm}^{\rm
cr}\right)$] and far from the dynamo threshold
($\lambda \propto {\rm Rm}^{1/2}$). The influence
of the Golitsyn spectrum on the small-scale
dynamo instability is minor, and this spectrum of
magnetic fluctuations requires a finite
correlation time of the random velocity field.

\bigskip

We are grateful to Alexander Schekochihin for
stimulating discussions who initiated our study
of small-scale dynamo with the Golitsyn
spectrum. We acknowledge the NORDITA dynamo
program of 2011 for providing a stimulating
scientific atmosphere.

\appendix

\section{Growth rate of small-scale dynamo
instability in the vicinity of the dynamo
threshold}

Let us obtain the scaling for the growth rate of
the small-scale dynamo instability in the
vicinity of the dynamo threshold. In the range of
scales, $\ell_\nu/\ell_0 < r \ll
\ell_\eta/\ell_0$, the function $r W'/W$ is given
by
\begin{eqnarray}
{r W'\over W} = - {4 \over 9} \, {\rm Rm} \,
r^{4/3} , \label{AB1}
\end{eqnarray}
[see Eq.~(\ref{B38})], while in the range of
scales, $\ell_\eta/\ell_0 \ll r < 1$ the function
$rW'/W$ is
\begin{eqnarray}
{r W'\over W} = - {13\over 6} - \sqrt{13\over 12}
\, \tan\left(\sqrt{13\over 12} \, \ln r +
\varphi_0 \right) ,
\label{AB2}
\end{eqnarray}
[see Eq.~(\ref{B5})]. On the other hand, in the
range of scales $r \gg 1$ the function $rW'/W$ is
\begin{eqnarray}
{r W'\over W} = - 2 - \sqrt{\lambda} \, r - {1
\over 1 + \sqrt{\lambda} \, r} \approx
-\left(3+\lambda r^2 \right),
\label{AB3}
\end{eqnarray}
[see Eq.~(\ref{B40})], where we have taken into
account that in the vicinity of the dynamo
threshold $\lambda \to 0$ and $\lambda^{1/2} \, r
\ll 1$.

Matching of the functions $rW'/W$ determined by
Eqs.~(\ref{AB1}) and~(\ref{AB2}) at the point $r
= \ell_\eta/\ell_0$ yields the following
equation:
\begin{eqnarray}
\tan\left({\sqrt{39}\over 8} \, \ln {\rm Rm} -
\varphi_0 \right) = {31 \over 3\sqrt{39}}.
\label{AB4}
\end{eqnarray}
This equation determines the function
$\varphi_0({\rm Rm})$. Now we define the function
$\varphi_0^{\rm cr}=\varphi_0({\rm Rm}={\rm
Rm}^{\rm cr})$, where ${\rm Rm}^{\rm cr}$ is the
threshold for the excitation of the magnetic
fluctuations. It follows from this equation that
\begin{eqnarray}
\varphi_0 - \varphi_0^{\rm cr} ={\sqrt{39}\over
8} \, \ln\left({{\rm Rm} \over {\rm Rm}^{\rm
cr}}\right). \label{AB5}
\end{eqnarray}
Matching of the functions $rW'/W$ determined by
Eqs.~(\ref{AB2}) and~(\ref{AB3}) at the point $r
= 1$ yields
\begin{eqnarray}
\tan\varphi_0 = \sqrt{12\over 13} \,
\left({5\over 6}+\lambda \right). \label{AB6}
\end{eqnarray}
It follows from this equation that
\begin{eqnarray}
\lambda = {32 \over  3\sqrt{39}} \,
\left(\varphi_0 - \varphi_0^{\rm cr} \right) ,
\label{AB7}
\end{eqnarray}
where we have also taken into account that in the
vicinity of the dynamo threshold $\lambda \to 0$.
Combining Eqs.~(\ref{AB5}) and~(\ref{AB7}), we
obtain the following scalinf for the growth rate
of small-scale dynamo instability in the vicinity
of the dynamo threshold: $\lambda = (4/3)
\ln\left({\rm Rm} / {\rm Rm}^{\rm cr}\right)$.

\section{Tangling magnetic fluctuations in the
delta-correlated-in-time velocity field}

The technique of path integrals for the
delta-correlated-in-time velocity field allows us
to derive the equation for the second-order
correlation function, $ h_{ij} = \langle
b_{i}(t,{\bf x}) b_{j}(t,{\bf y}) \rangle$:
\begin{eqnarray}
{\partial h_{ij} \over \partial t} = [ \hat
L_{ik}({\bf x}) \delta_{js} + \hat L_{js}({\bf
y}) \delta_{ik} + \hat M_{ijks} ] h_{ks} +
I_{ij},
 \label{L14}
\end{eqnarray}
(see for details \cite{RK97}), where the
turbulent component of magnetic field is ${\bf
b}(t,{\bf x}) = {\bf H}(t,{\bf x}) - {\bf B}(t)$,
\begin{eqnarray}
\hat L_{ij} &=& {1 \over 3}\left(1 +{3 \over {\rm
Rm}}\right) \delta_{ij} {\partial \over
\partial x_p} \, {\partial \over
\partial x_p} ,
\label{L23} \\
{1 \over 2} \hat M_{ijks} &=& \delta_{ik}
\delta_{js} f_{mn} {\partial^{2} \over \partial
x_m \partial y_{n} } - \delta_{ik} {\partial
f_{mj} \over \partial y_{s}} {\partial \over
\partial x_m} - \delta_{js} {\partial f_{in}
\over \partial x_{k}} {\partial \over \partial
y_n} + {\partial^{2} f_{ij} \over \partial x_k
\partial y_{s} },
\label{L24} \\
I_{ij} &=& B_{k} B_{s} {\partial^{2} f_{ij} \over
\partial x_k \partial y_{s} },
\label{L35}
\end{eqnarray}
and $f_{mn} = \langle \tau u_{m}({\bf x})
u_{n}({\bf y}) \rangle$. Multiplying Eq.
(\ref{L14}) by $ r_{i} r_{j} / r^{2} $ we arrive
at the equation for the correlation function
$\tilde W(r,t)$:
\begin{eqnarray}
{\partial \tilde W(t,r) \over \partial t} = {1
\over m(r)} \tilde W'' + \mu(r) \tilde W' -
{\kappa(r) \over m(r)} \tilde W + I, \label{L10}
\end{eqnarray}
where $I=B^2\, \left(F'' + 4F'/r\right)/3$. In
the inertial range the source term is $I=52 B^2
r^{-2/3} /27$. This source term yields the
following scaling of the correlation function
$W(r) \propto r^{4/3}$ in the range of scales,
$\ell_\nu/\ell_0 < r \ll \ell_\eta/\ell_0$ . This
implies that the scaling of magnetic fluctuations
caused by the tangling of large-scale magnetic
field by the delta-correlated-in-time velocity
field coincides with the scaling of turbulent
magnetic diffusion $F(r) \propto r^{4/3}$. In
Fourier space this corresponds to the $k^{-7/3}$
spectrum of the magnetic fluctuations. This
implies that the Golitsyn spectrum, $k^{-11/3}$,
of magnetic fluctuations cannot be described in
terms of the delta-correlated-in-time velocity
field.

\end{document}